\documentclass[
 amsmath,amssymb,
 aps,
prl,
]{revtex4-1}

\usepackage{graphicx}
\usepackage{dcolumn}
\usepackage{bm}

\usepackage{lipsum}

\begin{document}


\title{Ballistic Properties of Highly Stretchable Graphene Kirigami Pyramid}

\author{Alirio Moura and Douglas S. Galvao}
 \email{galvao@ifi.unicamp.br}
\affiliation{%
 Applied Physics Department and Center for Computational Engineering and Sciences, State University of Campinas (UNICAMP), Campinas, SP, 13083-969, Brazil.
}%

\author{Pedro Alves da Silva Autreto}
\affiliation{
Center of Human and Natural Sciences Federal University of ABC, Santo Andre, SP, Brazil.
}%

\date{\today}

\begin{abstract}
 Graphene kirigami (patterned cuts) can be an effective way to improve some of the graphene mechanical and electronic properties. In this work, we report the first study of the mechanical and ballistic behavior of single and multilayered graphene pyramid kirigami (GKP). We have carried out fully atomistic reactive molecular dynamics simulations. GPK presents a unique kinetic energy absorption due to its topology that creates multi-steps dissipation mechanisms, which block crack propagation. Our results show that even having significantly less mass, GKP can outperform graphene structures with similar dimensions in terms of absorbing kinetic energy.


\begin{description}
\item[PACs numbers: 62.25.−g, 61.48.De, 61.46.+w]
\end{description}
\end{abstract}

\pacs{Valid PACS appear here}
\maketitle


Graphene was the first two-dimensional (2D) material experimentally realized \cite{geim2007rise}. Its relatively simple and cheap fabrication method combined with its extraordinary electronic, thermal, and mechanical properties \cite{balandin2011thermal, neto2009electronic} have made graphene one of 'hottest' subject in materials science. It has been exploited in a large variety of applications, such as foams \cite{vinod2014low}, stretchable electronics \cite{kim2009large}, composites \cite{stankovich2006graphene}, etc. Besides its remarkable electronic properties, graphene has also mechanical ones, being the ultimate membrane (just a single atom thick). It has been experimentally demonstrated that graphene is several times more resilient to ballistic impact than steel \cite{lee2014dynamic} and with a unique petal-like fracture dynamics \cite{lee2014dynamic,bizao2018scale}.

However, in spite of its large stiffness and high strain values, graphene is very brittle, which limits its use for some applications \cite{kim2009large}. Graphene under strain will fracture immediately after its yield point, occasionally fragmenting into several parts. One way to overcome these problems is using patterned graphene nanomeshes, as proposed by Zhu et al. \cite{zhu2014extremely} based on \cite{qi2014atomistic} coarse-grained simulations. Even for elongation about 50\% the meshes can remain compliant \cite{zhu2014extremely}. 

Recently \cite{qi2014atomistic,castle2014making}, these ideas were extended using kirigami concepts, which combines folding and cutting to create patterns \cite{castle2014making}. Graphene kirigamis were experimentally realized by McEuen's group in 2015 \cite{blees2015graphene}. More recently, these concepts have been used in kirigami nets \cite{araujo2018finding}, kirigami topology \cite{chen2016topological,grosso2015bending,castro2018symmetry}, kirigami mechanics \cite{moshe2019kirigami,rafsanjani2017buckling,hanakata2018accelerated}, kirigami electronics \cite{mortazavi2017thermal}, and even in some applications, such batteries \cite{song2015kirigami} and nanocomposites \cite{shyu2015kirigami}.  

Kirigami structures have many advantages, the structural cuts can create new degrees of freedom (tensile and/or rotational), thus generating more flexible and elastic configurations than  their parent structures \cite{song2015kirigami}. Also, the cuts (shape, number, and dimensions) allow to engineer an almost infinite number of structures, and machine learning techniques have been used to design and/or optimize kirigami structures \cite{hanakata2018accelerated}.

However, in spite of a large number of kirigami structures their potential use as effective ballistic materials has not been yet fully investigated. In this work, using fully atomistic reactive molecular dynamics simulations (MD), we investigated the mechanical and ballistic properties of graphene kirigami pyramid (GKP) (see Figure 1) \cite{blees2015graphene}. It is the first time that the ballistic and mechanical properties of these structures are reported.

\begin{figure}
    \centering
    \includegraphics[width=12.0cm]{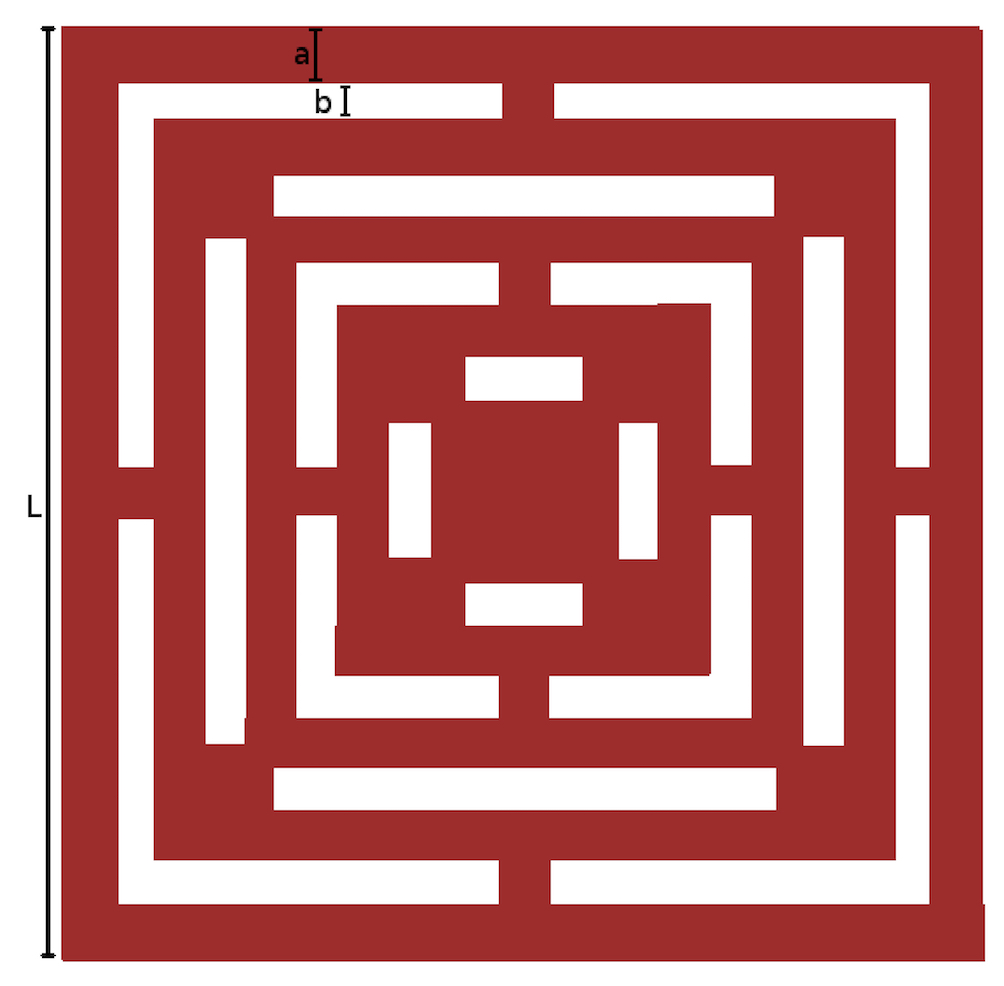}
    \caption{Schematic structure for graphene kirigami pyramid. \textbf{a} and \textbf{b} are structural parameters related to the graphene cuttings.}
    \label{fig1}
\end{figure}

All simulations were carried using the Large-scale Atomic-Molecular Massively Parallel Simulator(LAMMMPs) \cite{plimpton2007lammps} package. As we will investigate structural deformations beyond the plastic limit, it is necessary to use reactive force fields, i. e., force fields that can describe bond breaking and/or covalent bond formation. We chose ReaxFF \cite{van2001reaxff}. ReaxFF is a reactive force field developed by van Duin, Goddard III and co-workers for use in MD simulations \cite{van2001reaxff}. It allows simulations of many types of chemical reactions \cite{flores2009graphene,fluoro2013}. It is similar to standard non-reactive force fields, like MM3 \cite{allinger1989, van2001reaxff, van2003computational, chenoweth2008reaxff}, where the system energy is divided into partial energy contributions associated with, amongst others, valence angle bending and bond stretching, and non-bonded van der Waals and Coulomb interactions \cite{van2001reaxff, van2003computational, chenoweth2008reaxff}. One main difference is that ReaxFF can handle bond formation and dissociation (making/breaking bonds) as a function of bond order values. ReaxFF was parameterized against DFT calculations. Its average deviations from the predicted heats of formation values and the experimental ones are $2.8$ and $2.9$ kcal/mol, for non-conjugated and conjugated systems, respectively \cite{van2001reaxff, van2003computational, chenoweth2008reaxff}. 

In the present work, we used the Chenoweth et al. (2008) C/H/O ReaxFF parameterization  \cite{chenoweth2008reaxff} and all simulations were carried out with a time step of $0.1~fs$, with temperatures controlled through a Nos\'e-Hoover thermostat \cite{evans}. This methodology has been successfully used to study ballistic behavior of other carbon-based structures, such as graphene \cite{bizao2018scale}, nanotubes \cite{ozden2014unzipping} and nanoscrolls \cite{de2016carbon}. 

In order to obtain useful information regarding the  deformation and/or fracture dynamics from the MD simulations, we also calculated the virial stress tensor\cite{subramaniyan2008continuum, buellervonMises}, which can be defined as:

\begin{equation} 
\sigma_{ij}=\frac{\sum_{k}^{N}m_{k}v_{k_i}v_{k_j}}{V}+\frac{\sum_{k}^{N}r_{k_i}\cdot f_{k_j}}{V},
\end{equation}
where N is the number of atoms, $V$ is the volume, $m$ the mass of the atom, $v$ is the velocity, $r$ is the position and $f$ the force acting on the atom.
Stress-strain curves were obtained considering the relation between the uni-axial component of stress tensor along a specific direction, namely $\sigma_{ii}$, and 
the strain defined as a dimensionless quantity which is the ratio between deformation along the considered direction and the length on the same direction\cite{buellervonMises}
\begin{equation}
\varepsilon_i = \frac{\Delta Li}{Li},
\end{equation}
where $i=1,2$ or $3$. Using this quantity it is also useful to define the Young's Modulus, 
$Y=\sigma_{ii}/\varepsilon_i$, 
and the Poisson's ratio, which is the negative ratio between a transverse and an axial strain: 

\begin{equation}
\nu =-\frac{d\varepsilon_i}{d\varepsilon_j},
\end{equation} 

where $i\neq j$. We also calculated a quantity which is related to the second deviatoric stress invariant, also known as {\it von Mises stress}\cite{buellervonMises}, and defined as:

\begin{equation}
\sigma_{vm}=\sqrt{\frac{\left(\sigma_{11}-\sigma_{22}\right)^2+\left(\sigma_{22}-\sigma_{33}\right)^2+\left(\sigma_{11}-\sigma_{33}\right)^2
+6\left(\sigma_{12}^{2}+\sigma_{23}^{2}+\sigma_{31}^{2}\right)}{2}},
\label{vonMises}
\end{equation}
Components $\sigma_{12}$, $\sigma_{23}$ and $\sigma_{31}$ are the so-called shear stresses.  von Mises stress provides very helpful information on fracturing processes because, by calculating this quantity for each timestep, it is possible to visualize the time evolution and how the stress is spatially accumulated and/or dissipated. This methodology was successfully used to investigate the mechanical failure of carbon-based nanostructures such as graphene, carbon nanotubes \cite{dos2012unzipping}
and also silicon nanostructures\cite{buellervonMises}.

In Figure \ref{fig1} we show a schematic view of GKP based on a specific cutting pattern of a graphene sheet. \textbf{a} and \textbf{b} are structural parameters related to the graphene cutting. We tested many \textbf{a} and \textbf{b} values and the general trends are the same. The results presented here are for  \textbf{a}=8.5 and \textbf{b}=7.2 \AA, and L= 150 \AA.

We investigated GKP mechanical behavior under two regimes; under a static loading applied to the normal direction in the GKP central part. The protocol is quasi-statically pull up the GKP central square (see Figure \ref{fig1}) and then to allow the system to relax, and the von Miss stress is calculated.  During the simulations, the GKP outer border atoms are not allowed to move out-of-plane, but there freely to move in-plane. This protocol condition mimics a large graphene membrane. Another way of holding the sheet in-plane is to completely freeze the border atom position like stretching a tennis racket net, but this makes the material too stiff unless an impractically large (computational cost-prohibitive) graphene sheet is used. The loading process is continued up to limit where GPK fractures (mechanical failure). 

The second considered regime was a dynamical one, where a van der Waals solid sphere particle projectile with a radius of 15 \AA~ and mass= 15000 u.a. is ballistically shot against the GKP. We considered single and multi-layered (up to 5 layers) GKP. For comparison reasons graphene structures under the same conditions (same kinetic energy and area values) were also considered. The impact velocities were in the range of $50-100$ m/s.

\begin{figure}
    \centering
    \includegraphics[width=12.0cm]{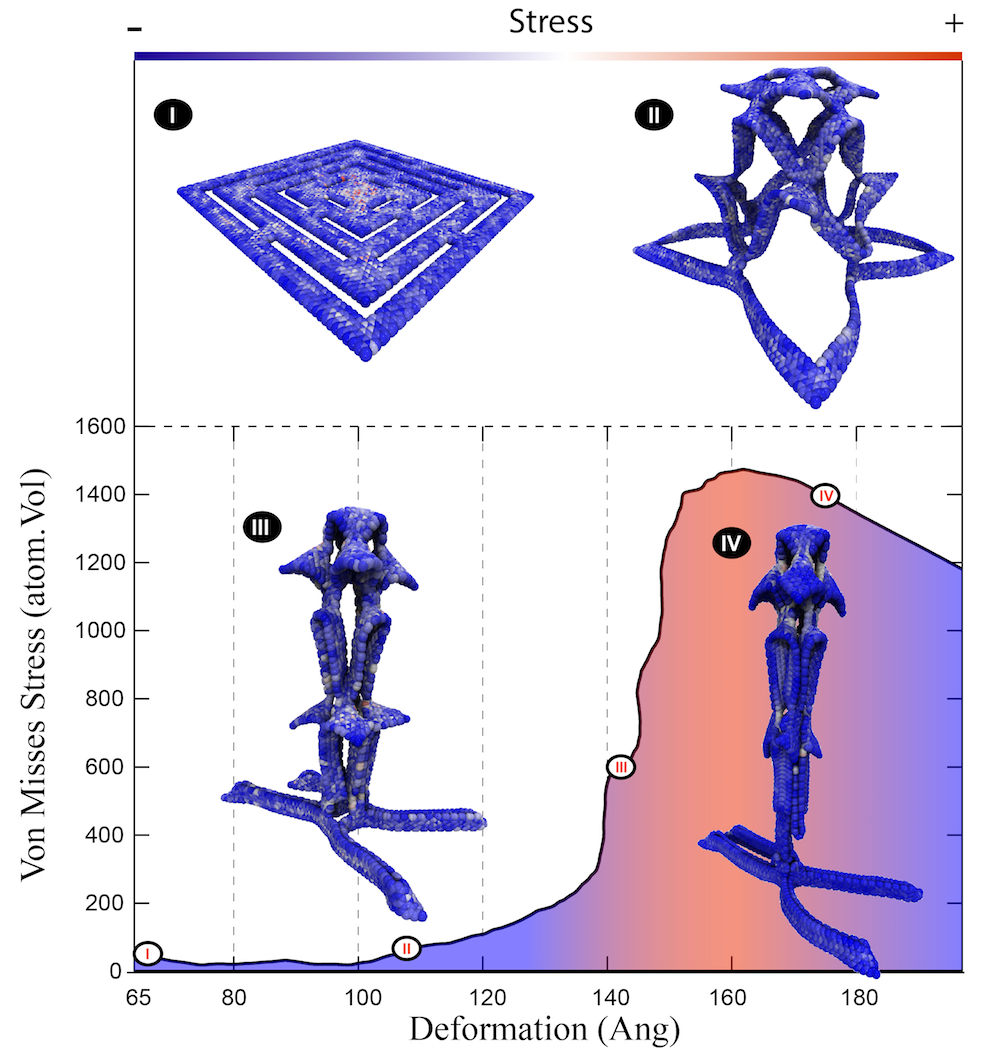}
    \caption{von Mises stress values as a function of structural deformations during the static loading. It is also shown representative configurations of the points indicated in the curve. The colors represent the stress regimes, from low (blue) to high (red).}
    \label{fig2}
\end{figure}

In Figure \ref{fig2} we present the von Mises stress values as a function of structural deformations under static loading. From this Figure we can identify 3 different regimes; during the first one, the force necessary to pull up the GKP central square is close of zero (I in Figure \ref{fig2}). After around $100$ \AA, we observe a quasi-linear region, then followed by an abrupt stress value increase ($140$ \AA (II in Figure \ref{fig2})). From there on the material reaches its limit of elastic deformation and partially breaks (III in Figure \ref{fig2}). After the fractures, as expected, there is a decrease in the stress values ((IV in Figure \ref{fig2}). The weak structural regions that always break first are the small bridges connecting the square rings that form the pyramid. These results show that kirigami topology significantly increase graphene stretchability, it is possible to stretch it almost 100\%~ without fracture. This is suggestive that this topology could result in a structure with good ballistic applications, which indeed was validated by the impact MD simulations, discussed below.

\begin{figure}
\centering
\includegraphics[width=7.5cm]{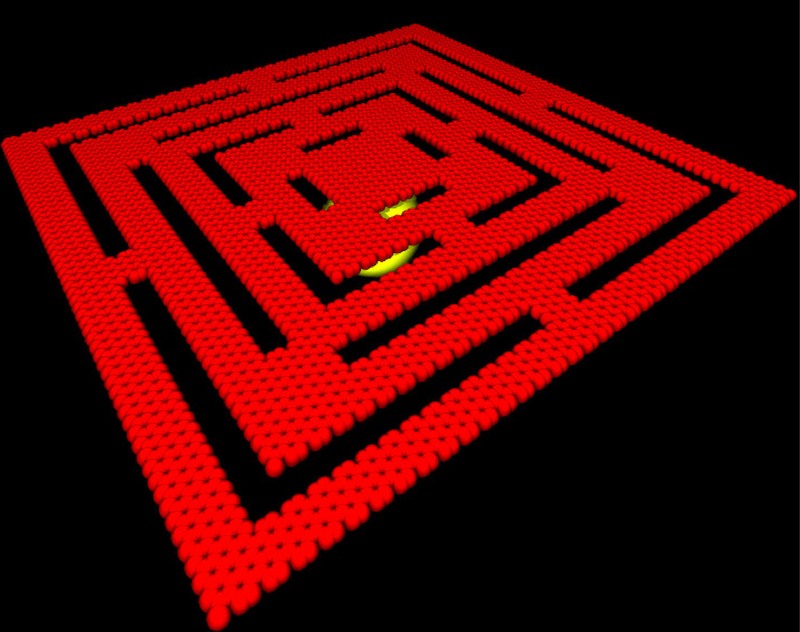}
\hspace{1em}
\includegraphics[width=7.5cm]{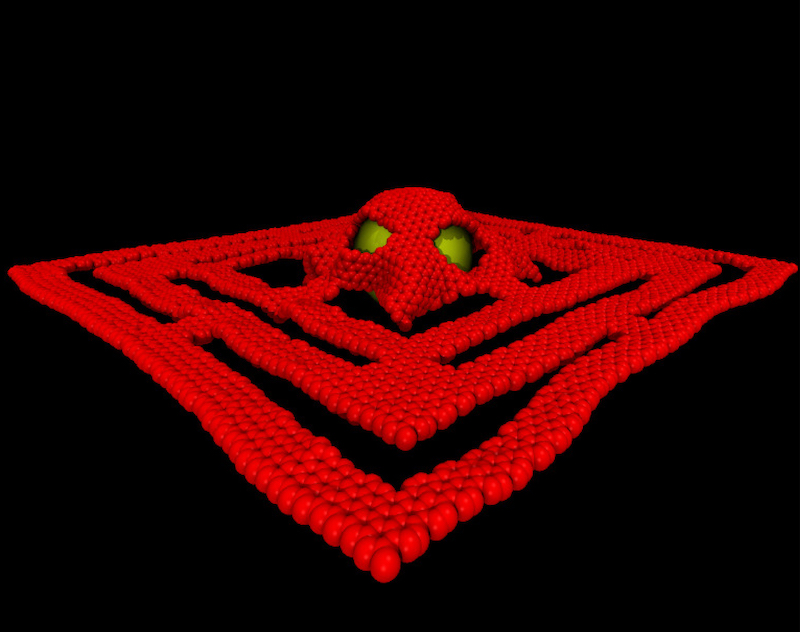}
\hspace{1em}
\includegraphics[width=7.5cm]{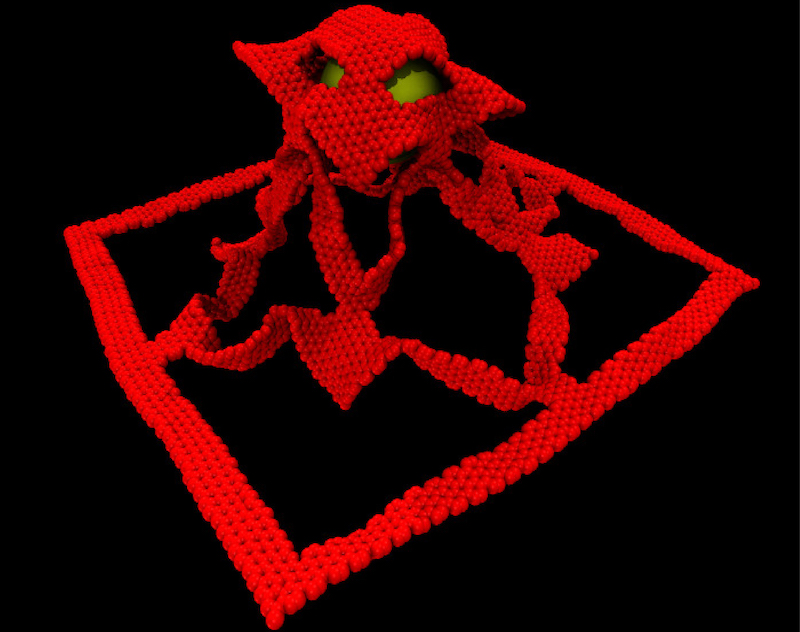}
\hspace{1em}
\includegraphics[width=7.5cm]{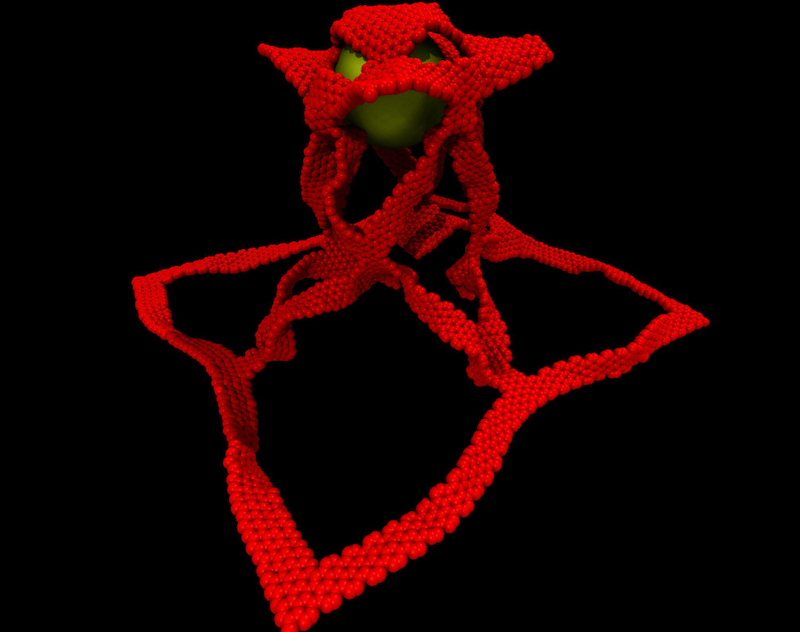}
\caption{Representative MD snapshots showing different stages of ballistically impacted single-layer graphene kirigami pyramid. Results for velocity impact of $50$ m/s.}
\label{fig3}
\end{figure}

In Figure \ref{fig3} we present representative MD snapshots for the case of a single-layer GKP at different stages, from just before impact up to completely stopping the projectile. As we can see from this Figure, similarly to the static loading case the GKP undergoes extensive deformations (including torsions). For the velocity case of $50$ m/s the GKP can stop the project without fracture (see video 1 in Supplementary Material \cite{supplementary}), while for the corresponding case, the graphene structure is perforated, with extensive structural damage. These differences are amplified, when the number of layers is increased (see videos 2 and 3 in Supplementary Material \cite{supplementary} for the bilayered structures).

\begin{figure}
    \centering
    \includegraphics[width=12.0cm]{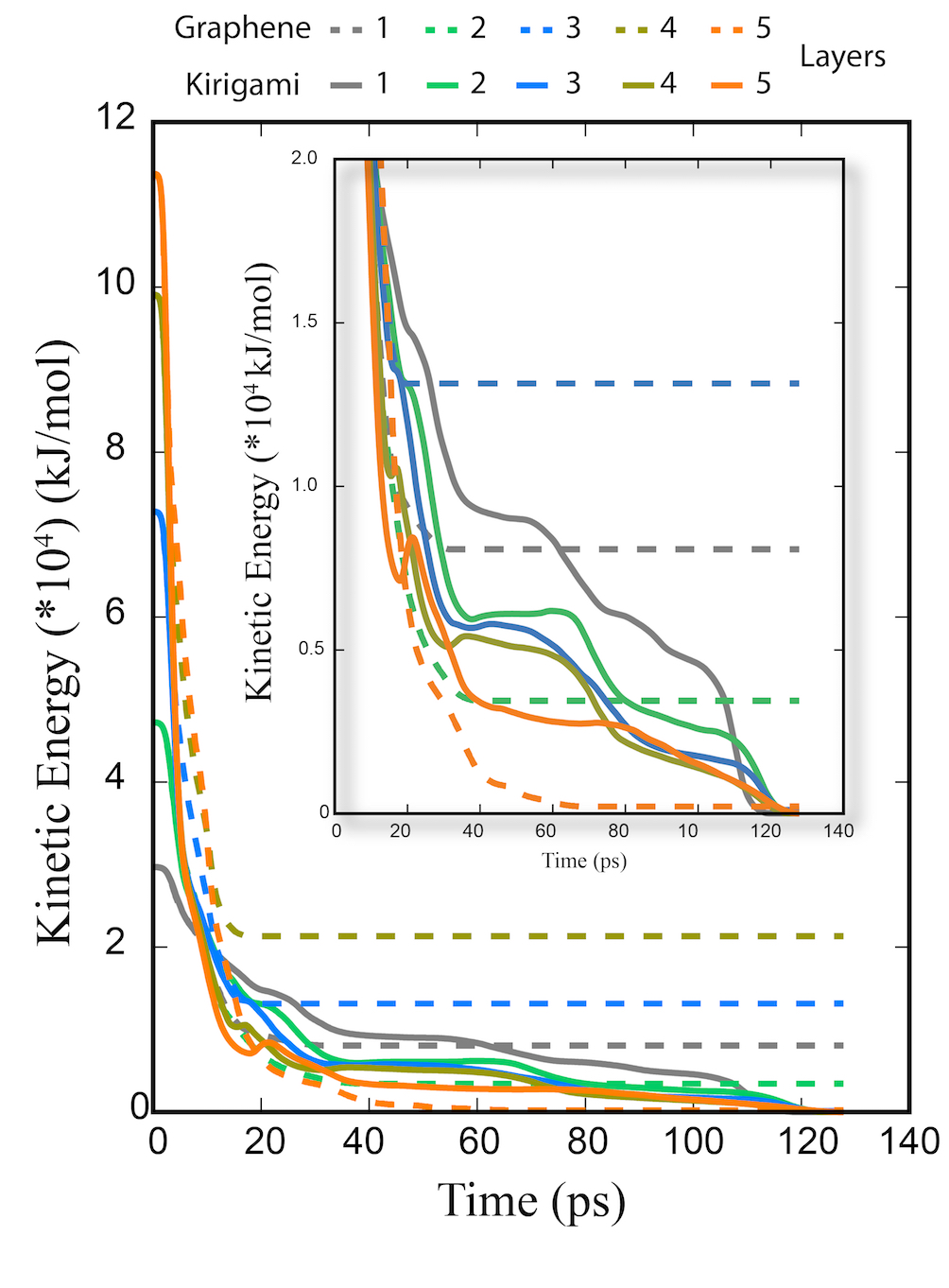}
    \caption{Projectile kinetic energy as a function of simulation time for different numbers of graphene and graphene kirigami pyramid layers. In the inset is shown a zoomed region.}
    \label{fig4}
\end{figure}

In Figure \ref{fig4} we present the projectile kinetic energy as a function of simulation time for multi-layered graphene and GKP. We present representative cases where the initial kinetic energy is below the GKP fracture threshold. As we can see from Figure \ref{fig4}, while for all cases of projectile kinetic energies considered here, the graphene structures are perforated (kinetic energy values become constant), the corresponding GKP ones can stop the projectile without fracture. Another interesting result, it is the existence of plateau-like structures present in Figure \ref{fig4} and this occur only for GKP. These plateau-like regimes indicate that the projectile can move with almost constant velocity during short periods of time and are indicative that the kirigami energy-absorption mechanisms occur in different stages and involving different parts of the structure. This can be better evidenced in videos 2 (bilayer GKP) and 3 (bilayer graphene) in the Supplementary Material \cite{supplementary}). It is remarkable that structures with the same lateral dimensions but lighter (GKP contains just 68\% of the total mass of their parent graphene structures) can absorb much more kinetic energy. This can be attributed to pure topological effects, the cuts create an efficient and larger number of dissipation channels, which avoids that crack propagation could result in total structural failures \cite{bizao2018scale}.  


In summary, we present the first study of the ballistic behavior of single and multi-layer graphene kirigami pyramid (GKP). Our results show that even containing less mass, GKP can outperform corresponding graphene structures of similar dimensions. GPK presents a unique kinetic energy-absorption due to their topology that create multi-steps dissipation mechanisms, which avoid crack propagation, making the structures more resilient to fractures. We hope the present study can stimulate further study along these lines. 

\bibliography{biblio}

\begin{thebibliography}{36}%
\makeatletter
\providecommand \@ifxundefined [1]{%
 \@ifx{#1\undefined}
}%
\providecommand \@ifnum [1]{%
 \ifnum #1\expandafter \@firstoftwo
 \else \expandafter \@secondoftwo
 \fi
}%
\providecommand \@ifx [1]{%
 \ifx #1\expandafter \@firstoftwo
 \else \expandafter \@secondoftwo
 \fi
}%
\providecommand \natexlab [1]{#1}%
\providecommand \enquote  [1]{``#1''}%
\providecommand \bibnamefont  [1]{#1}%
\providecommand \bibfnamefont [1]{#1}%
\providecommand \citenamefont [1]{#1}%
\providecommand \href@noop [0]{\@secondoftwo}%
\providecommand \href [0]{\begingroup \@sanitize@url \@href}%
\providecommand \@href[1]{\@@startlink{#1}\@@href}%
\providecommand \@@href[1]{\endgroup#1\@@endlink}%
\providecommand \@sanitize@url [0]{\catcode `\\12\catcode `\$12\catcode
  `\&12\catcode `\#12\catcode `\^12\catcode `\_12\catcode `\%12\relax}%
\providecommand \@@startlink[1]{}%
\providecommand \@@endlink[0]{}%
\providecommand \url  [0]{\begingroup\@sanitize@url \@url }%
\providecommand \@url [1]{\endgroup\@href {#1}{\urlprefix }}%
\providecommand \urlprefix  [0]{URL }%
\providecommand \Eprint [0]{\href }%
\providecommand \doibase [0]{http://dx.doi.org/}%
\providecommand \selectlanguage [0]{\@gobble}%
\providecommand \bibinfo  [0]{\@secondoftwo}%
\providecommand \bibfield  [0]{\@secondoftwo}%
\providecommand \translation [1]{[#1]}%
\providecommand \BibitemOpen [0]{}%
\providecommand \bibitemStop [0]{}%
\providecommand \bibitemNoStop [0]{.\EOS\space}%
\providecommand \EOS [0]{\spacefactor3000\relax}%
\providecommand \BibitemShut  [1]{\csname bibitem#1\endcsname}%
\let\auto@bib@innerbib\@empty
\bibitem [{\citenamefont {Geim}\ and\ \citenamefont
  {Novoselov}(2007)}]{geim2007rise}%
  \BibitemOpen
  \bibfield  {author} {\bibinfo {author} {\bibfnamefont {A.~K.}\ \bibnamefont
  {Geim}}\ and\ \bibinfo {author} {\bibfnamefont {K.~S.}\ \bibnamefont
  {Novoselov}},\ }\href@noop {} {\bibfield  {journal} {\bibinfo  {journal}
  {Nature materials}\ }\textbf {\bibinfo {volume} {6}},\ \bibinfo {pages} {183}
  (\bibinfo {year} {2007})}\BibitemShut {NoStop}%
\bibitem [{\citenamefont {Balandin}(2011)}]{balandin2011thermal}%
  \BibitemOpen
  \bibfield  {author} {\bibinfo {author} {\bibfnamefont {A.~A.}\ \bibnamefont
  {Balandin}},\ }\href@noop {} {\bibfield  {journal} {\bibinfo  {journal}
  {Nature materials}\ }\textbf {\bibinfo {volume} {10}},\ \bibinfo {pages}
  {569} (\bibinfo {year} {2011})}\BibitemShut {NoStop}%
\bibitem [{\citenamefont {Neto}\ \emph {et~al.}(2009)\citenamefont {Neto},
  \citenamefont {Guinea}, \citenamefont {Peres}, \citenamefont {Novoselov},\
  and\ \citenamefont {Geim}}]{neto2009electronic}%
  \BibitemOpen
  \bibfield  {author} {\bibinfo {author} {\bibfnamefont {A.~C.}\ \bibnamefont
  {Neto}}, \bibinfo {author} {\bibfnamefont {F.}~\bibnamefont {Guinea}},
  \bibinfo {author} {\bibfnamefont {N.~M.}\ \bibnamefont {Peres}}, \bibinfo
  {author} {\bibfnamefont {K.~S.}\ \bibnamefont {Novoselov}}, \ and\ \bibinfo
  {author} {\bibfnamefont {A.~K.}\ \bibnamefont {Geim}},\ }\href@noop {}
  {\bibfield  {journal} {\bibinfo  {journal} {Reviews of modern physics}\
  }\textbf {\bibinfo {volume} {81}},\ \bibinfo {pages} {109} (\bibinfo {year}
  {2009})}\BibitemShut {NoStop}%
\bibitem [{\citenamefont {Vinod}\ \emph {et~al.}(2014)\citenamefont {Vinod},
  \citenamefont {Tiwary}, \citenamefont {da~Silva~Autreto}, \citenamefont
  {Taha-Tijerina}, \citenamefont {Ozden}, \citenamefont {Chipara},
  \citenamefont {Vajtai}, \citenamefont {Galvao}, \citenamefont {Narayanan},\
  and\ \citenamefont {Ajayan}}]{vinod2014low}%
  \BibitemOpen
  \bibfield  {author} {\bibinfo {author} {\bibfnamefont {S.}~\bibnamefont
  {Vinod}}, \bibinfo {author} {\bibfnamefont {C.~S.}\ \bibnamefont {Tiwary}},
  \bibinfo {author} {\bibfnamefont {P.~A.}\ \bibnamefont {da~Silva~Autreto}},
  \bibinfo {author} {\bibfnamefont {J.}~\bibnamefont {Taha-Tijerina}}, \bibinfo
  {author} {\bibfnamefont {S.}~\bibnamefont {Ozden}}, \bibinfo {author}
  {\bibfnamefont {A.~C.}\ \bibnamefont {Chipara}}, \bibinfo {author}
  {\bibfnamefont {R.}~\bibnamefont {Vajtai}}, \bibinfo {author} {\bibfnamefont
  {D.~S.}\ \bibnamefont {Galvao}}, \bibinfo {author} {\bibfnamefont {T.~N.}\
  \bibnamefont {Narayanan}}, \ and\ \bibinfo {author} {\bibfnamefont {P.~M.}\
  \bibnamefont {Ajayan}},\ }\href@noop {} {\bibfield  {journal} {\bibinfo
  {journal} {Nature communications}\ }\textbf {\bibinfo {volume} {5}} (\bibinfo
  {year} {2014})}\BibitemShut {NoStop}%
\bibitem [{\citenamefont {Kim}\ \emph {et~al.}(2009)\citenamefont {Kim},
  \citenamefont {Zhao}, \citenamefont {Jang}, \citenamefont {Lee},
  \citenamefont {Kim}, \citenamefont {Kim}, \citenamefont {Ahn}, \citenamefont
  {Kim}, \citenamefont {Choi},\ and\ \citenamefont {Hong}}]{kim2009large}%
  \BibitemOpen
  \bibfield  {author} {\bibinfo {author} {\bibfnamefont {K.~S.}\ \bibnamefont
  {Kim}}, \bibinfo {author} {\bibfnamefont {Y.}~\bibnamefont {Zhao}}, \bibinfo
  {author} {\bibfnamefont {H.}~\bibnamefont {Jang}}, \bibinfo {author}
  {\bibfnamefont {S.~Y.}\ \bibnamefont {Lee}}, \bibinfo {author} {\bibfnamefont
  {J.~M.}\ \bibnamefont {Kim}}, \bibinfo {author} {\bibfnamefont {K.~S.}\
  \bibnamefont {Kim}}, \bibinfo {author} {\bibfnamefont {J.-H.}\ \bibnamefont
  {Ahn}}, \bibinfo {author} {\bibfnamefont {P.}~\bibnamefont {Kim}}, \bibinfo
  {author} {\bibfnamefont {J.-Y.}\ \bibnamefont {Choi}}, \ and\ \bibinfo
  {author} {\bibfnamefont {B.~H.}\ \bibnamefont {Hong}},\ }\href@noop {}
  {\bibfield  {journal} {\bibinfo  {journal} {Nature}\ }\textbf {\bibinfo
  {volume} {457}},\ \bibinfo {pages} {706} (\bibinfo {year}
  {2009})}\BibitemShut {NoStop}%
\bibitem [{\citenamefont {Stankovich}\ \emph {et~al.}(2006)\citenamefont
  {Stankovich}, \citenamefont {Dikin}, \citenamefont {Dommett}, \citenamefont
  {Kohlhaas}, \citenamefont {Zimney}, \citenamefont {Stach}, \citenamefont
  {Piner}, \citenamefont {Nguyen},\ and\ \citenamefont
  {Ruoff}}]{stankovich2006graphene}%
  \BibitemOpen
  \bibfield  {author} {\bibinfo {author} {\bibfnamefont {S.}~\bibnamefont
  {Stankovich}}, \bibinfo {author} {\bibfnamefont {D.~A.}\ \bibnamefont
  {Dikin}}, \bibinfo {author} {\bibfnamefont {G.~H.}\ \bibnamefont {Dommett}},
  \bibinfo {author} {\bibfnamefont {K.~M.}\ \bibnamefont {Kohlhaas}}, \bibinfo
  {author} {\bibfnamefont {E.~J.}\ \bibnamefont {Zimney}}, \bibinfo {author}
  {\bibfnamefont {E.~A.}\ \bibnamefont {Stach}}, \bibinfo {author}
  {\bibfnamefont {R.~D.}\ \bibnamefont {Piner}}, \bibinfo {author}
  {\bibfnamefont {S.~T.}\ \bibnamefont {Nguyen}}, \ and\ \bibinfo {author}
  {\bibfnamefont {R.~S.}\ \bibnamefont {Ruoff}},\ }\href@noop {} {\bibfield
  {journal} {\bibinfo  {journal} {nature}\ }\textbf {\bibinfo {volume} {442}},\
  \bibinfo {pages} {282} (\bibinfo {year} {2006})}\BibitemShut {NoStop}%
\bibitem [{\citenamefont {Lee}\ \emph {et~al.}(2014)\citenamefont {Lee},
  \citenamefont {Loya}, \citenamefont {Lou},\ and\ \citenamefont
  {Thomas}}]{lee2014dynamic}%
  \BibitemOpen
  \bibfield  {author} {\bibinfo {author} {\bibfnamefont {J.-H.}\ \bibnamefont
  {Lee}}, \bibinfo {author} {\bibfnamefont {P.~E.}\ \bibnamefont {Loya}},
  \bibinfo {author} {\bibfnamefont {J.}~\bibnamefont {Lou}}, \ and\ \bibinfo
  {author} {\bibfnamefont {E.~L.}\ \bibnamefont {Thomas}},\ }\href@noop {}
  {\bibfield  {journal} {\bibinfo  {journal} {Science}\ }\textbf {\bibinfo
  {volume} {346}},\ \bibinfo {pages} {1092} (\bibinfo {year}
  {2014})}\BibitemShut {NoStop}%
\bibitem [{\citenamefont {Bizao}\ \emph {et~al.}(2018)\citenamefont {Bizao},
  \citenamefont {Machado}, \citenamefont {de~Sousa}, \citenamefont {Pugno},\
  and\ \citenamefont {Galvao}}]{bizao2018scale}%
  \BibitemOpen
  \bibfield  {author} {\bibinfo {author} {\bibfnamefont {R.~A.}\ \bibnamefont
  {Bizao}}, \bibinfo {author} {\bibfnamefont {L.~D.}\ \bibnamefont {Machado}},
  \bibinfo {author} {\bibfnamefont {J.~M.}\ \bibnamefont {de~Sousa}}, \bibinfo
  {author} {\bibfnamefont {N.~M.}\ \bibnamefont {Pugno}}, \ and\ \bibinfo
  {author} {\bibfnamefont {D.~S.}\ \bibnamefont {Galvao}},\ }\href@noop {}
  {\bibfield  {journal} {\bibinfo  {journal} {Scientific reports}\ }\textbf
  {\bibinfo {volume} {8}},\ \bibinfo {pages} {6750} (\bibinfo {year}
  {2018})}\BibitemShut {NoStop}%
\bibitem [{\citenamefont {Zhu}\ \emph {et~al.}(2014)\citenamefont {Zhu},
  \citenamefont {Huang},\ and\ \citenamefont {Li}}]{zhu2014extremely}%
  \BibitemOpen
  \bibfield  {author} {\bibinfo {author} {\bibfnamefont {S.}~\bibnamefont
  {Zhu}}, \bibinfo {author} {\bibfnamefont {Y.}~\bibnamefont {Huang}}, \ and\
  \bibinfo {author} {\bibfnamefont {T.}~\bibnamefont {Li}},\ }\href@noop {}
  {\bibfield  {journal} {\bibinfo  {journal} {Applied Physics Letters}\
  }\textbf {\bibinfo {volume} {104}},\ \bibinfo {pages} {173103} (\bibinfo
  {year} {2014})}\BibitemShut {NoStop}%
\bibitem [{\citenamefont {Qi}\ \emph {et~al.}(2014)\citenamefont {Qi},
  \citenamefont {Campbell},\ and\ \citenamefont {Park}}]{qi2014atomistic}%
  \BibitemOpen
  \bibfield  {author} {\bibinfo {author} {\bibfnamefont {Z.}~\bibnamefont
  {Qi}}, \bibinfo {author} {\bibfnamefont {D.~K.}\ \bibnamefont {Campbell}}, \
  and\ \bibinfo {author} {\bibfnamefont {H.~S.}\ \bibnamefont {Park}},\
  }\href@noop {} {\bibfield  {journal} {\bibinfo  {journal} {Physical Review
  B}\ }\textbf {\bibinfo {volume} {90}},\ \bibinfo {pages} {245437} (\bibinfo
  {year} {2014})}\BibitemShut {NoStop}%
\bibitem [{\citenamefont {Castle}\ \emph {et~al.}(2014)\citenamefont {Castle},
  \citenamefont {Cho}, \citenamefont {Gong}, \citenamefont {Jung},
  \citenamefont {Sussman}, \citenamefont {Yang},\ and\ \citenamefont
  {Kamien}}]{castle2014making}%
  \BibitemOpen
  \bibfield  {author} {\bibinfo {author} {\bibfnamefont {T.}~\bibnamefont
  {Castle}}, \bibinfo {author} {\bibfnamefont {Y.}~\bibnamefont {Cho}},
  \bibinfo {author} {\bibfnamefont {X.}~\bibnamefont {Gong}}, \bibinfo {author}
  {\bibfnamefont {E.}~\bibnamefont {Jung}}, \bibinfo {author} {\bibfnamefont
  {D.~M.}\ \bibnamefont {Sussman}}, \bibinfo {author} {\bibfnamefont
  {S.}~\bibnamefont {Yang}}, \ and\ \bibinfo {author} {\bibfnamefont {R.~D.}\
  \bibnamefont {Kamien}},\ }\href@noop {} {\bibfield  {journal} {\bibinfo
  {journal} {Physical review letters}\ }\textbf {\bibinfo {volume} {113}},\
  \bibinfo {pages} {245502} (\bibinfo {year} {2014})}\BibitemShut {NoStop}%
\bibitem [{\citenamefont {Blees}\ \emph {et~al.}(2015)\citenamefont {Blees},
  \citenamefont {Barnard}, \citenamefont {Rose}, \citenamefont {Roberts},
  \citenamefont {McGill}, \citenamefont {Huang}, \citenamefont {Ruyack},
  \citenamefont {Kevek}, \citenamefont {Kobrin}, \citenamefont {Muller} \emph
  {et~al.}}]{blees2015graphene}%
  \BibitemOpen
  \bibfield  {author} {\bibinfo {author} {\bibfnamefont {M.~K.}\ \bibnamefont
  {Blees}}, \bibinfo {author} {\bibfnamefont {A.~W.}\ \bibnamefont {Barnard}},
  \bibinfo {author} {\bibfnamefont {P.~A.}\ \bibnamefont {Rose}}, \bibinfo
  {author} {\bibfnamefont {S.~P.}\ \bibnamefont {Roberts}}, \bibinfo {author}
  {\bibfnamefont {K.~L.}\ \bibnamefont {McGill}}, \bibinfo {author}
  {\bibfnamefont {P.~Y.}\ \bibnamefont {Huang}}, \bibinfo {author}
  {\bibfnamefont {A.~R.}\ \bibnamefont {Ruyack}}, \bibinfo {author}
  {\bibfnamefont {J.~W.}\ \bibnamefont {Kevek}}, \bibinfo {author}
  {\bibfnamefont {B.}~\bibnamefont {Kobrin}}, \bibinfo {author} {\bibfnamefont
  {D.~A.}\ \bibnamefont {Muller}},  \emph {et~al.},\ }\href@noop {} {\bibfield
  {journal} {\bibinfo  {journal} {Nature}\ }\textbf {\bibinfo {volume} {524}},\
  \bibinfo {pages} {204} (\bibinfo {year} {2015})}\BibitemShut {NoStop}%
\bibitem [{\citenamefont {Ara{\'u}jo}\ \emph {et~al.}(2018)\citenamefont
  {Ara{\'u}jo}, \citenamefont {da~Costa}, \citenamefont {Dorogovtsev},\ and\
  \citenamefont {Mendes}}]{araujo2018finding}%
  \BibitemOpen
  \bibfield  {author} {\bibinfo {author} {\bibfnamefont {N.}~\bibnamefont
  {Ara{\'u}jo}}, \bibinfo {author} {\bibfnamefont {R.}~\bibnamefont
  {da~Costa}}, \bibinfo {author} {\bibfnamefont {S.~N.}\ \bibnamefont
  {Dorogovtsev}}, \ and\ \bibinfo {author} {\bibfnamefont {J.}~\bibnamefont
  {Mendes}},\ }\href@noop {} {\bibfield  {journal} {\bibinfo  {journal}
  {Physical review letters}\ }\textbf {\bibinfo {volume} {120}},\ \bibinfo
  {pages} {188001} (\bibinfo {year} {2018})}\BibitemShut {NoStop}%
\bibitem [{\citenamefont {Chen}\ \emph {et~al.}(2016)\citenamefont {Chen},
  \citenamefont {Liu}, \citenamefont {Evans}, \citenamefont {Paulose},
  \citenamefont {Cohen}, \citenamefont {Vitelli},\ and\ \citenamefont
  {Santangelo}}]{chen2016topological}%
  \BibitemOpen
  \bibfield  {author} {\bibinfo {author} {\bibfnamefont {B.~G.-g.}\
  \bibnamefont {Chen}}, \bibinfo {author} {\bibfnamefont {B.}~\bibnamefont
  {Liu}}, \bibinfo {author} {\bibfnamefont {A.~A.}\ \bibnamefont {Evans}},
  \bibinfo {author} {\bibfnamefont {J.}~\bibnamefont {Paulose}}, \bibinfo
  {author} {\bibfnamefont {I.}~\bibnamefont {Cohen}}, \bibinfo {author}
  {\bibfnamefont {V.}~\bibnamefont {Vitelli}}, \ and\ \bibinfo {author}
  {\bibfnamefont {C.}~\bibnamefont {Santangelo}},\ }\href@noop {} {\bibfield
  {journal} {\bibinfo  {journal} {Physical review letters}\ }\textbf {\bibinfo
  {volume} {116}},\ \bibinfo {pages} {135501} (\bibinfo {year}
  {2016})}\BibitemShut {NoStop}%
\bibitem [{\citenamefont {Grosso}\ and\ \citenamefont
  {Mele}(2015)}]{grosso2015bending}%
  \BibitemOpen
  \bibfield  {author} {\bibinfo {author} {\bibfnamefont {B.~F.}\ \bibnamefont
  {Grosso}}\ and\ \bibinfo {author} {\bibfnamefont {E.}~\bibnamefont {Mele}},\
  }\href@noop {} {\bibfield  {journal} {\bibinfo  {journal} {Physical review
  letters}\ }\textbf {\bibinfo {volume} {115}},\ \bibinfo {pages} {195501}
  (\bibinfo {year} {2015})}\BibitemShut {NoStop}%
\bibitem [{\citenamefont {Castro}\ \emph {et~al.}(2018)\citenamefont {Castro},
  \citenamefont {Flachi}, \citenamefont {Ribeiro},\ and\ \citenamefont
  {Vitagliano}}]{castro2018symmetry}%
  \BibitemOpen
  \bibfield  {author} {\bibinfo {author} {\bibfnamefont {E.~V.}\ \bibnamefont
  {Castro}}, \bibinfo {author} {\bibfnamefont {A.}~\bibnamefont {Flachi}},
  \bibinfo {author} {\bibfnamefont {P.}~\bibnamefont {Ribeiro}}, \ and\
  \bibinfo {author} {\bibfnamefont {V.}~\bibnamefont {Vitagliano}},\
  }\href@noop {} {\bibfield  {journal} {\bibinfo  {journal} {Physical review
  letters}\ }\textbf {\bibinfo {volume} {121}},\ \bibinfo {pages} {221601}
  (\bibinfo {year} {2018})}\BibitemShut {NoStop}%
\bibitem [{\citenamefont {Moshe}\ \emph {et~al.}(2019)\citenamefont {Moshe},
  \citenamefont {Esposito}, \citenamefont {Shankar}, \citenamefont {Bircan},
  \citenamefont {Cohen}, \citenamefont {Nelson},\ and\ \citenamefont
  {Bowick}}]{moshe2019kirigami}%
  \BibitemOpen
  \bibfield  {author} {\bibinfo {author} {\bibfnamefont {M.}~\bibnamefont
  {Moshe}}, \bibinfo {author} {\bibfnamefont {E.}~\bibnamefont {Esposito}},
  \bibinfo {author} {\bibfnamefont {S.}~\bibnamefont {Shankar}}, \bibinfo
  {author} {\bibfnamefont {B.}~\bibnamefont {Bircan}}, \bibinfo {author}
  {\bibfnamefont {I.}~\bibnamefont {Cohen}}, \bibinfo {author} {\bibfnamefont
  {D.~R.}\ \bibnamefont {Nelson}}, \ and\ \bibinfo {author} {\bibfnamefont
  {M.~J.}\ \bibnamefont {Bowick}},\ }\href@noop {} {\bibfield  {journal}
  {\bibinfo  {journal} {Physical review letters}\ }\textbf {\bibinfo {volume}
  {122}},\ \bibinfo {pages} {048001} (\bibinfo {year} {2019})}\BibitemShut
  {NoStop}%
\bibitem [{\citenamefont {Rafsanjani}\ and\ \citenamefont
  {Bertoldi}(2017)}]{rafsanjani2017buckling}%
  \BibitemOpen
  \bibfield  {author} {\bibinfo {author} {\bibfnamefont {A.}~\bibnamefont
  {Rafsanjani}}\ and\ \bibinfo {author} {\bibfnamefont {K.}~\bibnamefont
  {Bertoldi}},\ }\href@noop {} {\bibfield  {journal} {\bibinfo  {journal}
  {Physical Review Letters}\ }\textbf {\bibinfo {volume} {118}},\ \bibinfo
  {pages} {084301} (\bibinfo {year} {2017})}\BibitemShut {NoStop}%
\bibitem [{\citenamefont {Hanakata}\ \emph {et~al.}(2018)\citenamefont
  {Hanakata}, \citenamefont {Cubuk}, \citenamefont {Campbell},\ and\
  \citenamefont {Park}}]{hanakata2018accelerated}%
  \BibitemOpen
  \bibfield  {author} {\bibinfo {author} {\bibfnamefont {P.~Z.}\ \bibnamefont
  {Hanakata}}, \bibinfo {author} {\bibfnamefont {E.~D.}\ \bibnamefont {Cubuk}},
  \bibinfo {author} {\bibfnamefont {D.~K.}\ \bibnamefont {Campbell}}, \ and\
  \bibinfo {author} {\bibfnamefont {H.~S.}\ \bibnamefont {Park}},\ }\href@noop
  {} {\bibfield  {journal} {\bibinfo  {journal} {Physical review letters}\
  }\textbf {\bibinfo {volume} {121}},\ \bibinfo {pages} {255304} (\bibinfo
  {year} {2018})}\BibitemShut {NoStop}%
\bibitem [{\citenamefont {Mortazavi}\ \emph {et~al.}(2017)\citenamefont
  {Mortazavi}, \citenamefont {Lherbier}, \citenamefont {Fan}, \citenamefont
  {Harju}, \citenamefont {Rabczuk},\ and\ \citenamefont
  {Charlier}}]{mortazavi2017thermal}%
  \BibitemOpen
  \bibfield  {author} {\bibinfo {author} {\bibfnamefont {B.}~\bibnamefont
  {Mortazavi}}, \bibinfo {author} {\bibfnamefont {A.}~\bibnamefont {Lherbier}},
  \bibinfo {author} {\bibfnamefont {Z.}~\bibnamefont {Fan}}, \bibinfo {author}
  {\bibfnamefont {A.}~\bibnamefont {Harju}}, \bibinfo {author} {\bibfnamefont
  {T.}~\bibnamefont {Rabczuk}}, \ and\ \bibinfo {author} {\bibfnamefont
  {J.-C.}\ \bibnamefont {Charlier}},\ }\href@noop {} {\bibfield  {journal}
  {\bibinfo  {journal} {Nanoscale}\ }\textbf {\bibinfo {volume} {9}},\ \bibinfo
  {pages} {16329} (\bibinfo {year} {2017})}\BibitemShut {NoStop}%
\bibitem [{\citenamefont {Song}\ \emph {et~al.}(2015)\citenamefont {Song},
  \citenamefont {Wang}, \citenamefont {Lv}, \citenamefont {An}, \citenamefont
  {Liang}, \citenamefont {Ma}, \citenamefont {He}, \citenamefont {Zheng},
  \citenamefont {Huang}, \citenamefont {Yu} \emph {et~al.}}]{song2015kirigami}%
  \BibitemOpen
  \bibfield  {author} {\bibinfo {author} {\bibfnamefont {Z.}~\bibnamefont
  {Song}}, \bibinfo {author} {\bibfnamefont {X.}~\bibnamefont {Wang}}, \bibinfo
  {author} {\bibfnamefont {C.}~\bibnamefont {Lv}}, \bibinfo {author}
  {\bibfnamefont {Y.}~\bibnamefont {An}}, \bibinfo {author} {\bibfnamefont
  {M.}~\bibnamefont {Liang}}, \bibinfo {author} {\bibfnamefont
  {T.}~\bibnamefont {Ma}}, \bibinfo {author} {\bibfnamefont {D.}~\bibnamefont
  {He}}, \bibinfo {author} {\bibfnamefont {Y.-J.}\ \bibnamefont {Zheng}},
  \bibinfo {author} {\bibfnamefont {S.-Q.}\ \bibnamefont {Huang}}, \bibinfo
  {author} {\bibfnamefont {H.}~\bibnamefont {Yu}},  \emph {et~al.},\
  }\href@noop {} {\bibfield  {journal} {\bibinfo  {journal} {Scientific
  reports}\ }\textbf {\bibinfo {volume} {5}} (\bibinfo {year}
  {2015})}\BibitemShut {NoStop}%
\bibitem [{\citenamefont {Shyu}\ \emph {et~al.}(2015)\citenamefont {Shyu},
  \citenamefont {Damasceno}, \citenamefont {Dodd}, \citenamefont {Lamoureux},
  \citenamefont {Xu}, \citenamefont {Shlian}, \citenamefont {Shtein},
  \citenamefont {Glotzer},\ and\ \citenamefont {Kotov}}]{shyu2015kirigami}%
  \BibitemOpen
  \bibfield  {author} {\bibinfo {author} {\bibfnamefont {T.~C.}\ \bibnamefont
  {Shyu}}, \bibinfo {author} {\bibfnamefont {P.~F.}\ \bibnamefont {Damasceno}},
  \bibinfo {author} {\bibfnamefont {P.~M.}\ \bibnamefont {Dodd}}, \bibinfo
  {author} {\bibfnamefont {A.}~\bibnamefont {Lamoureux}}, \bibinfo {author}
  {\bibfnamefont {L.}~\bibnamefont {Xu}}, \bibinfo {author} {\bibfnamefont
  {M.}~\bibnamefont {Shlian}}, \bibinfo {author} {\bibfnamefont
  {M.}~\bibnamefont {Shtein}}, \bibinfo {author} {\bibfnamefont {S.~C.}\
  \bibnamefont {Glotzer}}, \ and\ \bibinfo {author} {\bibfnamefont {N.~A.}\
  \bibnamefont {Kotov}},\ }\href@noop {} {\bibfield  {journal} {\bibinfo
  {journal} {Nature materials}\ }\textbf {\bibinfo {volume} {14}},\ \bibinfo
  {pages} {785} (\bibinfo {year} {2015})}\BibitemShut {NoStop}%
\bibitem [{\citenamefont {Plimpton}\ \emph {et~al.}(2007)\citenamefont
  {Plimpton}, \citenamefont {Crozier},\ and\ \citenamefont
  {Thompson}}]{plimpton2007lammps}%
  \BibitemOpen
  \bibfield  {author} {\bibinfo {author} {\bibfnamefont {S.}~\bibnamefont
  {Plimpton}}, \bibinfo {author} {\bibfnamefont {P.}~\bibnamefont {Crozier}}, \
  and\ \bibinfo {author} {\bibfnamefont {A.}~\bibnamefont {Thompson}},\
  }\href@noop {} {\bibfield  {journal} {\bibinfo  {journal} {Sandia National
  Laboratories}\ }\textbf {\bibinfo {volume} {18}} (\bibinfo {year}
  {2007})}\BibitemShut {NoStop}%
\bibitem [{\citenamefont {Van~Duin}\ \emph {et~al.}(2001)\citenamefont
  {Van~Duin}, \citenamefont {Dasgupta}, \citenamefont {Lorant},\ and\
  \citenamefont {Goddard}}]{van2001reaxff}%
  \BibitemOpen
  \bibfield  {author} {\bibinfo {author} {\bibfnamefont {A.~C.}\ \bibnamefont
  {Van~Duin}}, \bibinfo {author} {\bibfnamefont {S.}~\bibnamefont {Dasgupta}},
  \bibinfo {author} {\bibfnamefont {F.}~\bibnamefont {Lorant}}, \ and\ \bibinfo
  {author} {\bibfnamefont {W.~A.}\ \bibnamefont {Goddard}},\ }\href@noop {}
  {\bibfield  {journal} {\bibinfo  {journal} {The Journal of Physical Chemistry
  A}\ }\textbf {\bibinfo {volume} {105}},\ \bibinfo {pages} {9396} (\bibinfo
  {year} {2001})}\BibitemShut {NoStop}%
\bibitem [{\citenamefont {Flores}\ \emph {et~al.}(2009)\citenamefont {Flores},
  \citenamefont {Autreto}, \citenamefont {Legoas},\ and\ \citenamefont
  {Galvao}}]{flores2009graphene}%
  \BibitemOpen
  \bibfield  {author} {\bibinfo {author} {\bibfnamefont {M.~Z.}\ \bibnamefont
  {Flores}}, \bibinfo {author} {\bibfnamefont {P.~A.}\ \bibnamefont {Autreto}},
  \bibinfo {author} {\bibfnamefont {S.~B.}\ \bibnamefont {Legoas}}, \ and\
  \bibinfo {author} {\bibfnamefont {D.~S.}\ \bibnamefont {Galvao}},\
  }\href@noop {} {\bibfield  {journal} {\bibinfo  {journal} {Nanotechnology}\
  }\textbf {\bibinfo {volume} {20}},\ \bibinfo {pages} {465704} (\bibinfo
  {year} {2009})}\BibitemShut {NoStop}%
\bibitem [{\citenamefont {Paupitz}\ \emph {et~al.}(2013)\citenamefont
  {Paupitz}, \citenamefont {Autreto}, \citenamefont {Legoas}, \citenamefont
  {Srinivasan}, \citenamefont {van Duin},\ and\ \citenamefont
  {Galvao}}]{fluoro2013}%
  \BibitemOpen
  \bibfield  {author} {\bibinfo {author} {\bibfnamefont {R.}~\bibnamefont
  {Paupitz}}, \bibinfo {author} {\bibfnamefont {P.~A.~S.}\ \bibnamefont
  {Autreto}}, \bibinfo {author} {\bibfnamefont {S.~B.}\ \bibnamefont {Legoas}},
  \bibinfo {author} {\bibfnamefont {S.~G.}\ \bibnamefont {Srinivasan}},
  \bibinfo {author} {\bibfnamefont {T.}~\bibnamefont {van Duin}}, \ and\
  \bibinfo {author} {\bibfnamefont {D.~S.}\ \bibnamefont {Galvao}},\ }\href
  {\doibase 10.1088/0957-4484/24/3/035706} {\bibfield  {journal} {\bibinfo
  {journal} {Nanotechnology}\ }\textbf {\bibinfo {volume} {24}} (\bibinfo
  {year} {2013}),\ 10.1088/0957-4484/24/3/035706}\BibitemShut {NoStop}%
\bibitem [{\citenamefont {Allinger}\ \emph {et~al.}(1989)\citenamefont
  {Allinger}, \citenamefont {Yuh},\ and\ \citenamefont {Lii}}]{allinger1989}%
  \BibitemOpen
  \bibfield  {author} {\bibinfo {author} {\bibfnamefont {N.~L.}\ \bibnamefont
  {Allinger}}, \bibinfo {author} {\bibfnamefont {Y.~H.}\ \bibnamefont {Yuh}}, \
  and\ \bibinfo {author} {\bibfnamefont {J.~H.}\ \bibnamefont {Lii}},\
  }\href@noop {} {\bibfield  {journal} {\bibinfo  {journal} {Journal of the
  American Chemical Society}\ }\textbf {\bibinfo {volume} {111}},\ \bibinfo
  {pages} {8551} (\bibinfo {year} {1989})}\BibitemShut {NoStop}%
\bibitem [{\citenamefont {van Duin}\ and\ \citenamefont
  {Damst{\'e}}(2003)}]{van2003computational}%
  \BibitemOpen
  \bibfield  {author} {\bibinfo {author} {\bibfnamefont {A.~C.}\ \bibnamefont
  {van Duin}}\ and\ \bibinfo {author} {\bibfnamefont {J.~S.~S.}\ \bibnamefont
  {Damst{\'e}}},\ }\href@noop {} {\bibfield  {journal} {\bibinfo  {journal}
  {Organic Geochemistry}\ }\textbf {\bibinfo {volume} {34}},\ \bibinfo {pages}
  {515} (\bibinfo {year} {2003})}\BibitemShut {NoStop}%
\bibitem [{\citenamefont {Chenoweth}\ \emph {et~al.}(2008)\citenamefont
  {Chenoweth}, \citenamefont {Van~Duin},\ and\ \citenamefont
  {Goddard}}]{chenoweth2008reaxff}%
  \BibitemOpen
  \bibfield  {author} {\bibinfo {author} {\bibfnamefont {K.}~\bibnamefont
  {Chenoweth}}, \bibinfo {author} {\bibfnamefont {A.~C.}\ \bibnamefont
  {Van~Duin}}, \ and\ \bibinfo {author} {\bibfnamefont {W.~A.}\ \bibnamefont
  {Goddard}},\ }\href@noop {} {\bibfield  {journal} {\bibinfo  {journal} {The
  Journal of Physical Chemistry A}\ }\textbf {\bibinfo {volume} {112}},\
  \bibinfo {pages} {1040} (\bibinfo {year} {2008})}\BibitemShut {NoStop}%
\bibitem [{\citenamefont {Evans}\ and\ \citenamefont {Holian}(1985)}]{evans}%
  \BibitemOpen
  \bibfield  {author} {\bibinfo {author} {\bibfnamefont {D.~J.}\ \bibnamefont
  {Evans}}\ and\ \bibinfo {author} {\bibfnamefont {B.~L.}\ \bibnamefont
  {Holian}},\ }\href@noop {} {\bibfield  {journal} {\bibinfo  {journal} {The
  Journal of chemical physics}\ }\textbf {\bibinfo {volume} {83}},\ \bibinfo
  {pages} {4069} (\bibinfo {year} {1985})}\BibitemShut {NoStop}%
\bibitem [{\citenamefont {Ozden}\ \emph {et~al.}(2014)\citenamefont {Ozden},
  \citenamefont {Autreto}, \citenamefont {Tiwary}, \citenamefont {Khatiwada},
  \citenamefont {Machado}, \citenamefont {Galvao}, \citenamefont {Vajtai},
  \citenamefont {Barrera},\ and\ \citenamefont
  {M.~Ajayan}}]{ozden2014unzipping}%
  \BibitemOpen
  \bibfield  {author} {\bibinfo {author} {\bibfnamefont {S.}~\bibnamefont
  {Ozden}}, \bibinfo {author} {\bibfnamefont {P.~A.}\ \bibnamefont {Autreto}},
  \bibinfo {author} {\bibfnamefont {C.~S.}\ \bibnamefont {Tiwary}}, \bibinfo
  {author} {\bibfnamefont {S.}~\bibnamefont {Khatiwada}}, \bibinfo {author}
  {\bibfnamefont {L.}~\bibnamefont {Machado}}, \bibinfo {author} {\bibfnamefont
  {D.~S.}\ \bibnamefont {Galvao}}, \bibinfo {author} {\bibfnamefont
  {R.}~\bibnamefont {Vajtai}}, \bibinfo {author} {\bibfnamefont {E.~V.}\
  \bibnamefont {Barrera}}, \ and\ \bibinfo {author} {\bibfnamefont
  {P.}~\bibnamefont {M.~Ajayan}},\ }\href@noop {} {\bibfield  {journal}
  {\bibinfo  {journal} {Nano letters}\ }\textbf {\bibinfo {volume} {14}},\
  \bibinfo {pages} {4131} (\bibinfo {year} {2014})}\BibitemShut {NoStop}%
\bibitem [{\citenamefont {de~Sousa}\ \emph {et~al.}(2016)\citenamefont
  {de~Sousa}, \citenamefont {Machado}, \citenamefont {Woellner}, \citenamefont
  {da~Silva~Autreto},\ and\ \citenamefont {Galvao}}]{de2016carbon}%
  \BibitemOpen
  \bibfield  {author} {\bibinfo {author} {\bibfnamefont {J.~M.}\ \bibnamefont
  {de~Sousa}}, \bibinfo {author} {\bibfnamefont {L.~D.}\ \bibnamefont
  {Machado}}, \bibinfo {author} {\bibfnamefont {C.~F.}\ \bibnamefont
  {Woellner}}, \bibinfo {author} {\bibfnamefont {P.~A.}\ \bibnamefont
  {da~Silva~Autreto}}, \ and\ \bibinfo {author} {\bibfnamefont {D.~S.}\
  \bibnamefont {Galvao}},\ }\href@noop {} {\bibfield  {journal} {\bibinfo
  {journal} {MRS Advances}\ ,\ \bibinfo {pages} {1}} (\bibinfo {year}
  {2016})}\BibitemShut {NoStop}%
\bibitem [{\citenamefont {Subramaniyan}\ and\ \citenamefont
  {Sun}(2008)}]{subramaniyan2008continuum}%
  \BibitemOpen
  \bibfield  {author} {\bibinfo {author} {\bibfnamefont {A.~K.}\ \bibnamefont
  {Subramaniyan}}\ and\ \bibinfo {author} {\bibfnamefont {C.}~\bibnamefont
  {Sun}},\ }\href@noop {} {\bibfield  {journal} {\bibinfo  {journal}
  {International Journal of Solids and Structures}\ }\textbf {\bibinfo {volume}
  {45}},\ \bibinfo {pages} {4340} (\bibinfo {year} {2008})}\BibitemShut
  {NoStop}%
\bibitem [{\citenamefont {Garcia}\ and\ \citenamefont
  {Buehler}(2010)}]{buellervonMises}%
  \BibitemOpen
  \bibfield  {author} {\bibinfo {author} {\bibfnamefont {A.~P.}\ \bibnamefont
  {Garcia}}\ and\ \bibinfo {author} {\bibfnamefont {M.~J.}\ \bibnamefont
  {Buehler}},\ }\href@noop {} {\bibfield  {journal} {\bibinfo  {journal}
  {Computational Materials Science}\ }\textbf {\bibinfo {volume} {48}},\
  \bibinfo {pages} {303} (\bibinfo {year} {2010})}\BibitemShut {NoStop}%
\bibitem [{\citenamefont {Dos~Santos}\ \emph {et~al.}(2012)\citenamefont
  {Dos~Santos}, \citenamefont {Perim}, \citenamefont {Autreto}, \citenamefont
  {Brunetto},\ and\ \citenamefont {Galvao}}]{dos2012unzipping}%
  \BibitemOpen
  \bibfield  {author} {\bibinfo {author} {\bibfnamefont {R.}~\bibnamefont
  {Dos~Santos}}, \bibinfo {author} {\bibfnamefont {E.}~\bibnamefont {Perim}},
  \bibinfo {author} {\bibfnamefont {P.}~\bibnamefont {Autreto}}, \bibinfo
  {author} {\bibfnamefont {G.}~\bibnamefont {Brunetto}}, \ and\ \bibinfo
  {author} {\bibfnamefont {D.}~\bibnamefont {Galvao}},\ }\href@noop {}
  {\bibfield  {journal} {\bibinfo  {journal} {Nanotechnology}\ }\textbf
  {\bibinfo {volume} {23}},\ \bibinfo {pages} {465702} (\bibinfo {year}
  {2012})}\BibitemShut {NoStop}%
\bibitem [{sup()}]{supplementary}%
  \BibitemOpen
  \href@noop {} {\bibinfo  {journal} {See Supplemental Material at LINK for
  additional details on ballistic test}\ }\BibitemShut {NoStop}%
\end{thebibliography}%

\begin{section}{ACKNOWLEDGMENTS}
This work was supported in part by the Brazilian Agencies, CAPES, CNPq and
  FAPESP. The authors thank the Center for Computational Engineering
  and Sciences at Unicamp for financial support through the
  FAPESP/CEPID Grant \#2013/08293-7. 
  \end{section}

\end{document}